\documentclass[10pt,twocolumn]{article} 
\usepackage{simpleConference}
\usepackage{times}
\usepackage{graphicx}
\usepackage{amssymb}
\usepackage{url,hyperref}
\usepackage{float} 

\begin{document}

\title{Analysis of a microcirculatory windkessel model using photoplethysmography with green light: A pilot study}

\author{ \href{https://orcid.org/0000-0002-0558-360X}{\hspace{1mm}Akio Tanaka} \\
\\
Tanaka Research Institute \\
Sagamihara, Kanagawa, Japan \\
\\
\\
contact.tanaka.giken@gmail.com  \\
}

\maketitle
\thispagestyle{empty}

\begin{abstract}
In this study, a vasomotion quantification method using a photoplethysmography prototype, which performs near-infrared spectroscopy in combination with green light, is proposed. A structure that suppresses the motion artifact and that is held by the eyeglasses on the back of the ear enables the relative concentration changes of total hemoglobin and pulse wave amplitude to be measured during exercise with and without the presence of wind impacting the face. We established a microcirculatory windkessel model including arteriovenous anastomoses estimated from the blood flow changes in the depth direction that were acquired using three wavelengths of light and reproduced the vasomotion on a computer. The values predicted by the model were in good agreement with the measured values. The extracted vasomotion can be used to understand autonomic control by the central nervous system.
\end{abstract}


\section{Introduction}
To understand dynamic behavior and/or simulate disease, research studies are underway to model {\it in vivo} phenomena and reproduce them on a computer. These models include a) sensors that monitor blood pressure or skin temperature (baroreceptors~\cite{R1} and skin thermoreceptors~\cite{R2}, respectively), b) elements of the central nervous system (cardiovascular~\cite{R1} and thermoregulatory center~\cite{R2}), c) autonomic nerves (sympathetic~\cite{R1,R2} and parasympathetic~\cite{R1}), d) blood flow in compliant vessels (windkessel models~\cite{R1,R3,R4,R5}), and e) vasomotion that modulates blood flow~\cite{R1,R2,R3,R4}. Verification of these models often uses invasive techniques and/or large equipment, such as catheters, magnetic resonance imaging, or laser-Doppler flowmetry~\cite{R2,R3}. Although many wearable devices for health monitoring have been developed~\cite{R6,R7}, it remains challenging to comprehensively collect highly accurate biophysical information. Therefore, multidisciplinary research is needed to understand and predict real-time physiological dynamics in real-life environments by merging modeling and wearable approaches. \par
A technique using near-infrared spectroscopy (NIRS) for photoplethysmography (PPG) has been reported for measuring both the arterial oxygen saturation ($SpO_2$) and heart rate ($HR$) from a pulsatile component ($AC$ signal) and also the relative concentration change of hemoglobin from a relatively slow varying component ($DC$ signal)~\cite{R8,R9,R10}. In general, most part of the $DC$ signal is contained in the capillaries and veins~\cite{R3}, and the $AC$ signal exists in the vessels through which blood flows before entering the capillaries. This technique enables to observe each change in blood volume in the microcirculation, where arterioles, capillaries, and venules represent a coordinated interplay~\cite{R11}. Usually, capillaries and venules passively respond to blood flow changes, which are modulated by vasomotor control of arterioles~\cite{R3}. \par
Furthermore, it has been reported that when a finger was exposed to heat, the infrared light $DC$ signal decreased, whereas that of the green light did not change. This reduction in $DC$ components was suggested to result from increased blood flow in the vascular bed, as infrared light has a large penetration depth, whereas green light has a shorter penetration depth~\cite{R12}. In contrast, skin blood flow changes largely due to thermoregulation~\cite{R13,R14}, and in the fingers or ears of humans and animals such as rabbits, there are many arteriovenous anastomoses (AVAs) that bypass arterioles and venules for thermoregulation~\cite{R14,R15,R16}. \par
To use the $DC$ and $AC$ signals of the PPG, motion artifacts (MAs), such as $DC$ level shifts due to the device moving from its original position or large and long fluctuations due to abnormal mechanical vibration, must be suppressed~\cite{R17,R18,R19,R20}. We have developed a prototype device with a PPG sensor that includes green light and is placed on the back of the ear and held by eyeglasses. As the head has little change in position with respect to the heart, the blood pressure in the head vasculature is almost stable. The integrity of the biophysical signals is also improved by adopting a structure that suppresses the generation of MAs. \par
In this study, experiments using a cycle ergometer, with and without wind, were performed on a subject. We created a microcirculatory windkessel model with AVAs from the acquired data using the NIRS processing technique, and demonstrated that vasomotion can be reproduced on a computer. Using the reproduced vasomotion, we discuss autonomic control during body temperature changes.

\section{Experimental setup}
\subsection{NIRS process with green light}

Using the $DC$ components of the red and infrared signals from the PPG, the relative change in total hemoglobin concentration ($\Delta [tHb]_{R, IR}$) can be obtained from the modified Beer–Lambert law as described in~\cite{R8}:

\begin{equation}
\Delta [tHb]_{R, IR} = e_1 \Delta A_R + e_2 \Delta A_{IR}, 
\label{eq1}
\end{equation}

\noindent
where $\Delta A_R$ and $\Delta A_{IR}$ are the changes from the start of the measurement of the $DC$ signals of red (R) or infrared (IR), respectively, and are expressed by $\Delta A_\lambda$ = ln$\frac {DC(0)} {DC(t)}$. The coefficients $e_1$ and $e_2$ can be obtained from the absorptivity coefficients of oxygenated hemoglobin ($HbO_2$) and reduced hemoglobin ($HHb$), at the red and infrared wavelengths. Similarly, $\Delta [tHb]_G$ at the green wavelength (G) can be obtained from

\begin{equation}
\Delta [tHb]_G \approx \frac {\Delta A_G} {\varepsilon_G}.
\label{eq2}
\end{equation}

\noindent
The wavelength of green light used here was 530 nm, and at this wavelength, the absorptivity coefficients of $HbO_2$ and $HHb$ are almost equal ($\varepsilon _G$); therefore, $\Delta [tHb]_G$ can be obtained from monochromatic light. Furthermore, an index of change in oxygenation $A_{Ox}$ that correlates with the tissue oxygenation index of NIRS~\cite{R8} can be obtained from

\begin{equation}
A_{Ox} = \Delta A_{IR} - \Delta A_R.
\label{eq3}
\end{equation}

\subsection{Prototype device}

\begin{figure}[!b]
\begin{center}
\includegraphics[width=8.5cm]{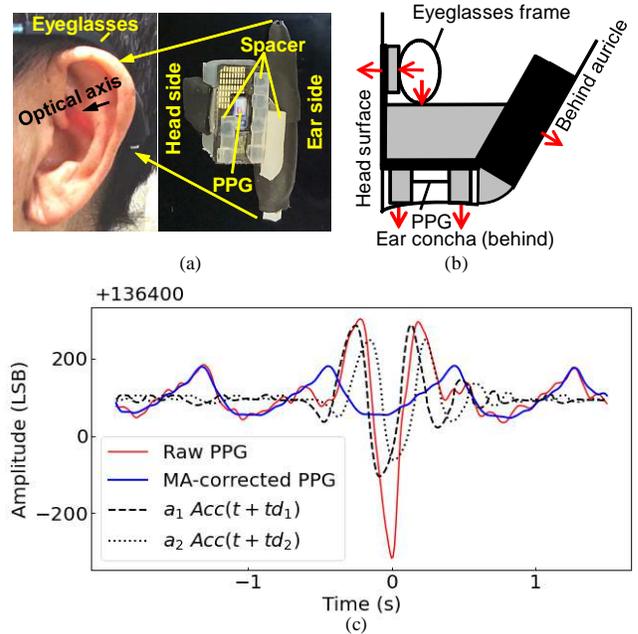}
\end{center}
\caption{\small (a) Developed prototype device and (b) corresponding cross-sectional structure. (c) Waveform obtained by wearing the device and jumping once.}
\label{fig:Fig1}
\end{figure}

The prototype device developed for the experiment and its cross-sectional structure are shown in Figs.~\ref{fig:Fig1}(a) and~\ref{fig:Fig1}(b), respectively. The device has a PPG sensor, accelerometer, 2.4-GHz radio, and battery. A reflectance PPG sensor (MAX30105, Maxim) with green, red, and infrared light-emitting diodes was used as the PPG sensor. The prototype device has a frame structure supported by the three skin surfaces behind the ear and a sensor head structure that avoids contact with MA-sensitive areas on the skin using spacers. Although it is preferable to attach the device to the same position each time, this can be achieved by checking the light emission from the PPG that has passed through the ear. The mass of the device was reduced to 4.8 g to reduce pressure on the skin.\par
Fig.~\ref{fig:Fig1}(c) shows the waveform obtained by wearing the device and jumping once. The MA components associated with the jump were superimposed on the raw PPG signal. These MA components were highly correlated with the acceleration signal $Acc(t)$ acquired by the accelerometer on the device. $Acc(t + td_1)$ and $Acc(t + td_1)$ in Fig.~\ref{fig:Fig1}(c) are the signals obtained by adding delays of $td_1$ and $td_2$, respectively, to the acceleration signal. By subtracting $a_1 Acc(t + td_1) + a_2 Acc(t + td_2)$ ($a_1$ and $a_2$ are coefficients) from the raw PPG signal, the MA components can be removed. The pulse waveforms of the subject are identifiable in the MA-corrected waveform. Owing to the structure that suppresses the MAs of the device, the spectra of the MAs are simplified into a composition of two acceleration waveforms, and $DC$ shifts can also be suppressed. Although it is possible to remove the MA components by correction, this was not done in this study because the MAs were sufficiently low.

\subsection{Experimental protocol}

Blood pressure has been reported to change by approximately -40\% even in healthy subjects when standing up~\cite{R21,R22}. During exercise, the cardiovascular system increases cardiac output and redistributes blood flow more efficiently to increase blood flow to active muscles~\cite{R23,R24}. In our preliminary experiments, $\Delta [tHb]$s ($\Delta [tHb]_G$ and $\Delta [tHb]_{R, IR}$) changed because of orthostatic stress when standing up and heat stress during exercise; therefore, the test protocol was examined based on these factors. \par

\begin{figure}[!b]
\begin{center}
\includegraphics[width=8.5cm]{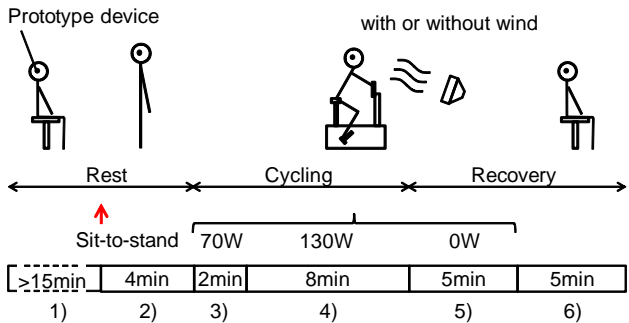}
\end{center}
\caption{\small Experimental protocol created by combining standing up, cool, and hot conditions.}
\label{fig:Fig2}
\end{figure}

The experimental protocol shown in Fig.~\ref{fig:Fig2} was created by combining standing up, cool, and hot conditions to excite changes in vasomotion. The protocol included the following states: 1) resting in a sitting position for at least 15 min, 2) standing up and remaining standing for 4 min, 3) exercising at 70 W using a cycle ergometer (AFB6013, ALINCO) for 2 min, 4) exercising at 130 W using a cycle ergometer for 8 min, 5) resting on the cycle ergometer for a recovery period of 5 min, and 6) recovering in a sitting position for 5 min. During steps 3 to 5, the experiments were conducted under two conditions: with and without wind on the face (the wind speed at the face position was 1.2 m/s). Room temperature was maintained between 27 \textdegree C and 28 \textdegree C. In this pilot study, changes in $\Delta [tHb]$s, $A_{Ox}$ and pulse waves were observed in one healthy participant when the above protocol was repeated multiple times.

\section{Experimental results} 
\subsection{Measured results}

\begin{figure}[!b]
\begin{center}
\includegraphics[width=8.5cm]{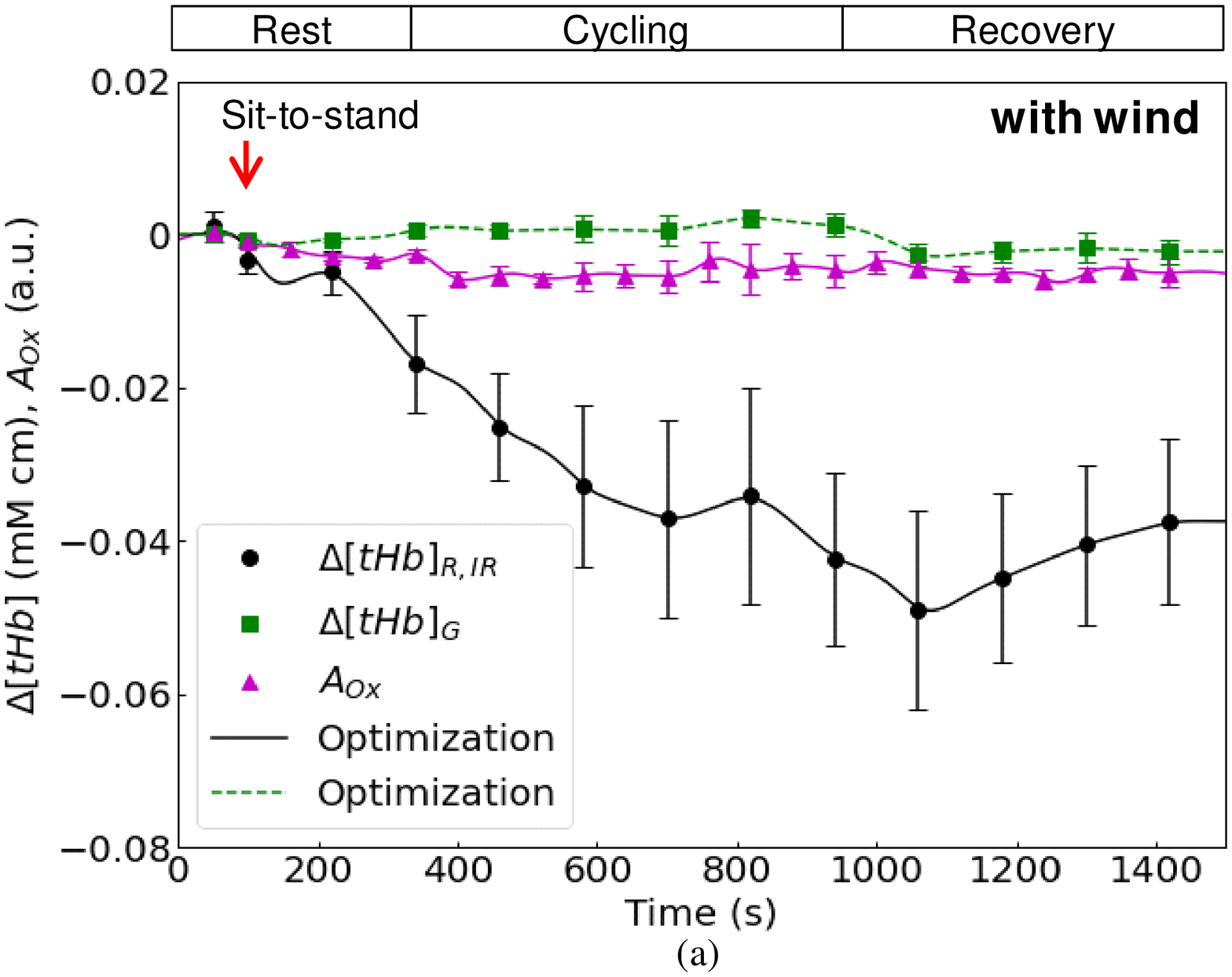}
\includegraphics[width=8.5cm]{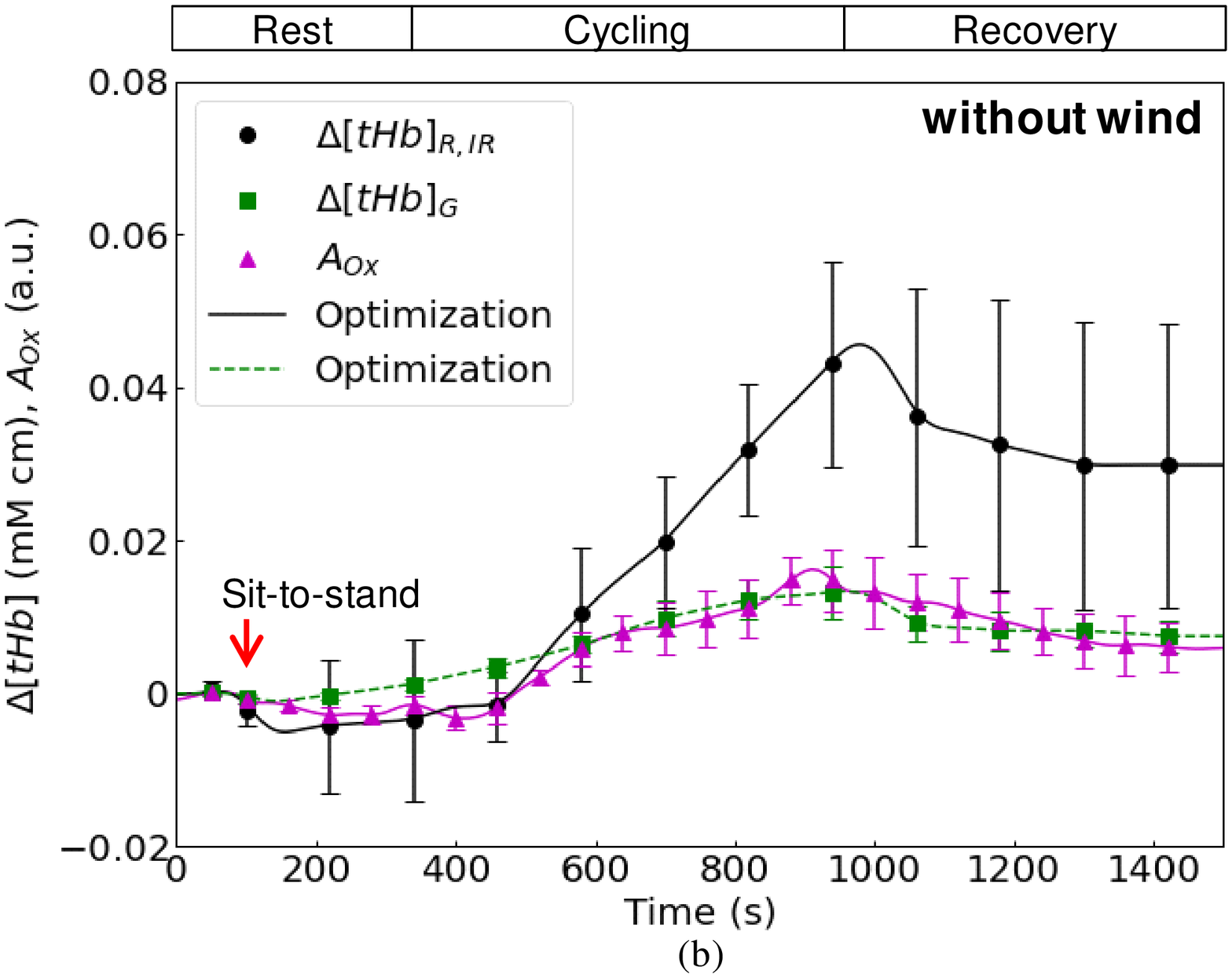}
\end{center}
\caption{\small Measured $\Delta [tHb]_{R, IR}$: relative changes in concentration of total hemoglobin from red and infrared light, $\Delta [tHb]_G$: from green light and $A_{Ox}$: index of changes in oxygenation (a) with wind and (b) without wind in the subject's face. Mean ± standard deviation (SD) are presented for each time point for (a) n = 5 and (b) n = 6. }
\label{fig:Fig3}
\end{figure}

The measured results with and without wind on the face are shown in Figs.~\ref{fig:Fig3}(a) and~\ref{fig:Fig3}(b), respectively. The time of 100 s that occurred before standing up was set to 0 s in the figure. $\Delta [tHb]_{R, IR}$ decreased significantly, whereas $\Delta [tHb]_G$ and $A_{Ox}$ did not change significantly in the presence of wind. $\Delta [tHb]$s and $A_{Ox}$ increased when cycling started and slowly decreased after cycling in the absence of wind.\par
\begin{figure}[!t]
\begin{center}
\includegraphics[width=8.5cm]{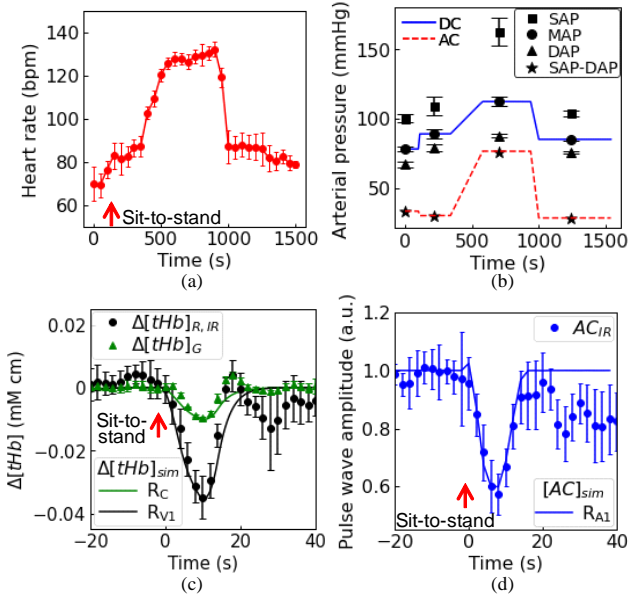}
\end{center}
\caption{
\small Measured (a) heart rate and (b) arterial pressure ($AP$): systolic ($SAP$), diastolic ($DAP$), and mean ($MAP$). (c) Measured waveforms of relative changes in concentration of total hemoglobin ($\Delta [tHb]$) and (d) pulse wave amplitude when standing up. Mean ± SD are presented for each time point for $AP$ (n = 4) and the others (n = 6).}
\label{fig:Fig4}
\end{figure}

The $HR$ extracted from the $AC$ PPG signal during the test and the change in arterial pressure ($AP$), measured with an automatic sphygmomanometer (ES-P302, TERUMO) attached to the upper arm, are shown in Figs.~\ref{fig:Fig4}(a) and~\ref{fig:Fig4}(b), respectively ($SAP$: systolic, $DAP$: diastolic, and $MAP$: mean arterial pressure). Because no major differences between the results with and without wind were observed, the blood pressure measurements were mixed and averaged, and only $HR$ without wind was plotted. Figs.~\ref{fig:Fig4}(c) and~\ref{fig:Fig4}(d) show the waveforms of $\Delta [tHb]$ and pulse wave amplitude ($PWA$) when standing up, respectively. $PWA$ is plotted as the root mean square (RMS) value of the $AC$ component of the infrared signal $AC(t)_{IR}$ normalized to the $AC(0)_{IR}$ value before standing up. Although only the infrared signal was plotted, the green and red signals exhibited almost identical changes. $\Delta [tHb]$s and $PWA$ decreased after standing and reached a minimum peak approximately 10 s later.

\subsection{Microcirculatory model with AVAs}

\begin{figure}[!t]
\begin{center}
\includegraphics[width=8.5cm]{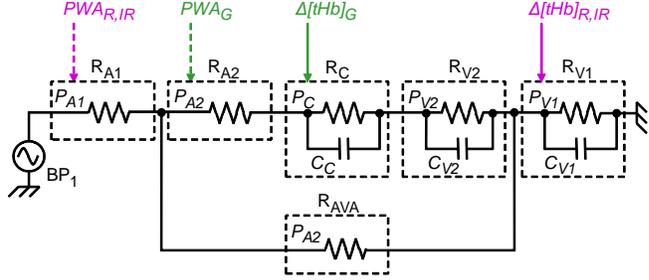}
\end{center}
\caption{\small Windkessel model of microcirculation including AVAs.}
\label{fig:Fig5}
\end{figure}

As shown in Fig.~\ref{fig:Fig3}(a), when wind was present, only $\Delta [tHb]_{R, IR}$ decreased significantly, whereas $\Delta [tHb]_{R, IR}$ and $\Delta [tHb]_G$ increased in the absence of wind (Fig.~\ref{fig:Fig3}(b)). It is necessary to assume factors for different behaviors between $\Delta [tHb]_{R, IR}$ and $\Delta [tHb]_G$. We created a windkessel model of microcirculation including AVAs, as shown in Fig.~\ref{fig:Fig5}, based on their existence in the ears~\cite{R3,R14,R15}, and studied whether this could explain the measured results. Here, the AVA compartment ($\rm{R_{AVA}}$) branched from the middle of the arteriolar compartments ($\rm{R_{A1}}$ and $\rm{R_{A2}}$) and joined the middle of the venule compartments ($\rm{R_{V2}}$ and $\rm{R_{V1}}$). The capillary compartment ($\rm{R_C}$) was located between $\rm{R_{A2}}$ and $\rm{R_{V2}}$. \par
In this study using this windkessel model, the following assumptions were made: i) green light measures $\Delta [tHb]_G$ in $\rm{R_C}$ and $PWA$ in $\rm{R_{A2}}$, and red and infrared lights measure $\Delta [tHb]_{R, IR}$ in $\rm{R_{V1}}$ and $PWA$ in $\rm{R_{A1}}$ (Fig.~\ref{fig:Fig5}); ii) $\rm{R_{A1}}$ and $\rm{R_{A2}}$ are under the same vasomotor control, and $\rm{R_{AVA}}$ is controlled by a different mechanism; iii) blood pressure measured in the upper arm during the test can be used for $DC$ and $AC$ signals of the blood pressure source ($\rm{BP_1}$); and iv) the windkessel model shown in Fig.~\ref{fig:Fig5} can be calibrated using the -40\% blood pressure variation observed when standing up.

\begin{table}[!t]
\centering
\small
\caption{\small Parameters used in windkessel model.} \label{tab:I}
\begin{tabular}{ccc}
\hline
Compartment & Pressure & Initial resistance (k$\Omega$) \\
\hline
$\rm BP_1$ & $BP_{DC} + BP_{AC}$ & \\
$\rm R_{A1}$ & $P_{A1} - P_{V1}$ & 65 \\
$\rm R_{A2}$ & $P_{A2} - P_{V1}$ & 65 \\
$\rm R_C$ & $P_C - P_{V1}$ & 25 \\
$\rm R_{V2}$ &$ P_{V2} - P_{V1}$ & 10 \\
$\rm R_{V1}$ & $P_{V1}$ & 10 \\
$\rm R_{AVA}$ & $P_{A2} - P_{V1}$ & 25 \\
\hline
\end{tabular}
\end{table}

\subsection{Models in the windkessel compartment}

Each vascular compartment uses a model that replaces changes in vessel volume, $V(t)$, with changes in resistance, $R(t)$, over time~\cite{R3}. A laminar flow ($\alpha = 2$) can be used for the model as follows:

\begin{equation}
\frac {R(t)} {R(0)} = \frac {1} {\delta} \left[\frac {V(0)} {V(t)}\right]^\alpha,
\label{eq4}
\end{equation}

\noindent
where $\delta$ is an index introduced for vasomotion in arterioles ($\delta_A$) and AVAs ($\delta_{AVA}$) in this study and is included in the denominator to provide a conductance function. Furthermore, volume ($V$) can be assumed to be related to blood pressure ($P$) in the compartment~\cite{R3} as

\begin{equation}
V \propto P^{1/\beta}.
\label{eq5}
\end{equation}

\noindent
In a previous study, $\beta \approx 1$ was used in a response with a long time constant of more than a few tens of seconds~\cite{R3}, and in another study, a sigmoid-shaped curve was used as this pressure-volume relation (linear near low pressure)~\cite{R25}. In this study, $\beta = 3$ was used in the $AC$ signals (i.e., they were rapid responses) when blood pressure was high, and $\beta = 1$ was used in other cases. The pressure $P$ in each compartment is the differential pressure between the pressure of each node and the pressure outside the vessels~\cite{R3}, for which $P_{V1}$ is used, as shown in Table~\ref{tab:I}. 
A SPICE circuit simulator was used for this analysis. The initial resistance for each compartment was chosen as shown in Table~\ref{tab:I}. Here, $R_{A20}:R_{C0}:R_{V20} = 0.65:0.25:0.1$ was chosen from~\cite{R3}, and $R_{A10} = R_{A20}$ and $R_{V10} = R_{V20}$ were chosen such that the simulated $\Delta [tHb]$ values matched the measured values. The absolute value of each initial resistance was determined such that SPICE converged. The values of capacitance ($C$) were determined such that the simulated transient responses matched those observed when standing up. The $DC$ and $AC$ signals of $\rm BP_1$ were modeled using the $MAP$ and pulse pressure ($SAP$ – $DAP$), respectively. The waveforms between the measured points were estimated from the change in $HR$, as shown in Fig.~\ref{fig:Fig4}(b).
From each $DC$ node voltage ($P_{DC}$) simulated by SPICE, $\Delta [tHb]$sim was calculated using Eq. (\ref{eq5}) as

\begin{equation}
\Delta [tHb]_{sim} = \gamma \left\{ \left[ \frac {P_{DC}(t)} {P_{DC}(0)} \right]^{1/\beta} -1 \right\} ,
\label{eq6}
\end{equation}

\noindent
where $\gamma$ is a calibration coefficient that fits the simulated $\Delta [tHb]$ to the actual measured value $(\gamma_G$ for $\Delta [tHb]_G$ and $\gamma_{R, IR}$ for $\Delta [tHb]_{R, IR}$). In the calibration, -40\% changes when standing up were applied to both the $DC$ and $AC$ $\rm BP_1$ signals. From each $AC$ voltage at the center of the compartment ($P_{AC}$) simulated by SPICE, each $PWA$ was calculated using Eqs. (\ref{eq4}) and (\ref{eq5}) as follows:
\begin{equation}
\left[\frac {AC(t)} {AC(0)}\right]_{sim} = \left[\frac {P_{AC}(t)} {P_{AC}(0)}\right] \left[\frac {P_{DC}(t)} {P_{DC}(0)} \right]^{\frac {1} {\beta} -1} {\delta _A}^{\frac {1} {\alpha}}.
\end{equation}

\subsection{Analysis of vasomotion}

\begin{figure}[!htb]
\begin{center}
\includegraphics[width=8.5cm]{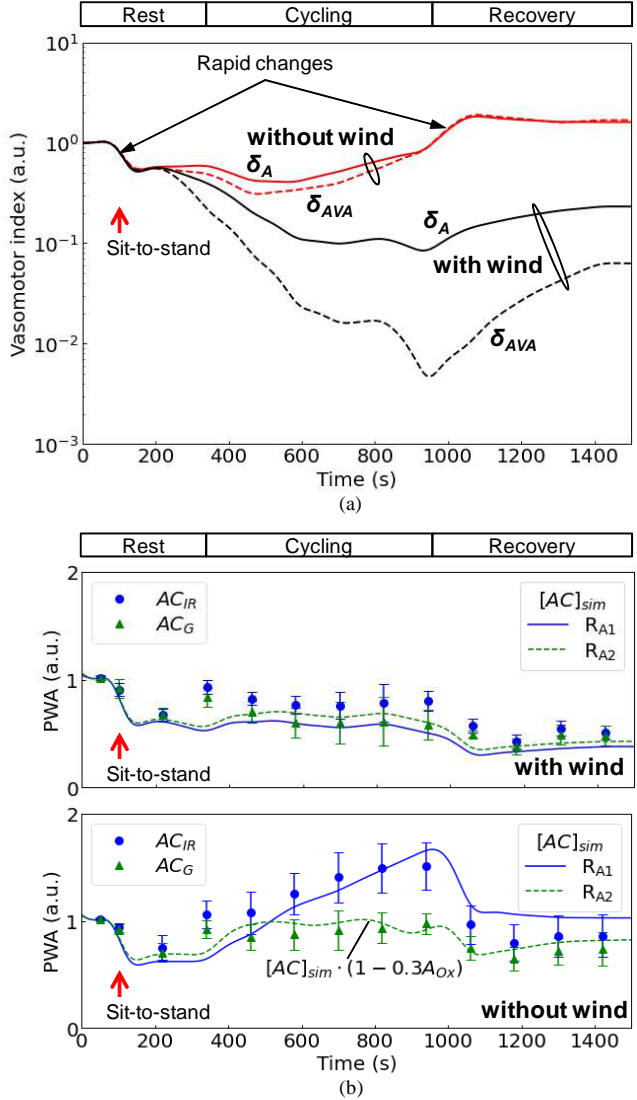}
\end{center}
\caption{\small (a) Vasomotor indices of arteriole ($\delta_A$) and AVA ($\delta_{AVA}$) compartments obtained by optimization. (b) Comparison of measured (infrared and green light) and simulated ($R_{A1}$ and $R_{A2}$) pulse wave amplitudes ($PWA$s) obtained using optimization results.}
\label{fig:Fig6}
\end{figure}

The solid lines in Fig.~\ref{fig:Fig4}(c) show the simulated results when standing after calibrating $\gamma_G$ and $\gamma_{R, IR}$. To reproduce the $PWA$ seen in Fig.~\ref{fig:Fig4}(d), $\delta_A$ and $\delta_{AVA}$ were set to 1. Vasomotion may not have occurred in the back of the ear when standing up. To analyze $\delta$s ($\delta_A$ and $\delta_{AVA}$) during the whole test using the calibrated model, they were set as two unknowns and optimized to minimize the difference between measured and simulated results. Although $\Delta [tHb]$ and $PWA$ are measured values, here we used $\Delta [tHb]$, which is determined only by the $DC$ characteristics and thus exhibits fewer errors. A SPICE optimizer was used for this search. The initial resistance of the $\rm R_{AVA}$ was set to 1/4 that of the capillary route $(R_{A2} + R_C + R_{V2})$ to obtain a solution. This is consistent with reports~\cite{R26} showing that approximately 80\% of the blood flows to AVAs at rest. The solid and dashed lines in Fig.~\ref{fig:Fig3} for $\Delta [tHb]$s are the results after optimization. $\delta$s values were obtained, as shown in Fig.~\ref{fig:Fig6}(a).

\section{Discussion}

In Fig.~\ref{fig:Fig6}(a) with wind, both $\delta_A$ and $\delta_{AVA}$ contracted significantly during cycling, particularly for $\delta_{AVA}$, which contracted to approximately 1/100 of its initial value. The central nervous system may detect a decrease in skin temperature and suppress blood flow to sites that are not necessary for heat dissipation to increase blood flow to active muscles~\cite{R2,R23,R24}. However, both $\delta_A$ and $\delta_{AVA}$ contracted slightly during cycling without wind and dilated during recovery. Because the increase in blood pressure is multiplied by $\delta$s, the blood volume $\Delta [tHb]$s increased, as shown in Fig.~\ref{fig:Fig3}(b), and heat flow out of the skin was promoted. The central nervous system may have received input from core body temperature~\cite{R2}, as $\delta$s dilated with a delay of several minutes. \par
Comparisons of the measured and simulated $PWA$s for infrared and green light obtained using the optimization results are shown in Fig.~\ref{fig:Fig6}(b). The solid and dashed lines represent the simulation results of $PWA$ in $\rm{R_{A1}}$ and $\rm{R_{A2}}$, respectively. Except for $\rm{R_{A2}}$ without wind, the actual measurements and simulations were in good agreement. The simulation result for $\rm{R_{A2}}$ without wind was multiplied by $(1 - 0.3 A_{Ox})$ to fit the measured value of green light. $A_{Ox}$ reflects the partial pressure of oxygen ($PaO_2$), and it increased without wind, as shown in Fig.~\ref{fig:Fig3}(b). Local contractions may have occurred in arterioles near capillaries~\cite{R27,R28} observed by green light owing to the increase in $PaO_2$. Thus, the windkessel model optimized using $\Delta [tHb]$s explained the $PWA$ measurement results well. \par
As shown in Fig.~\ref{fig:Fig3}(b), both $A_{Ox}$ and $\Delta [tHb]_G$ exhibited similar waveforms. AVAs may have decreased their vessel diameters with an increase in $PaO_2$, and this changed the $\Delta [tHb]_G$ value because the routes through the capillaries are sensitive to AVAs. This behavior contradicts the role of AVAs in increasing blood flow; however, as a result, both capillary and venous flows considerably increased. Although the increase in $A_{Ox}$ is considered to be due to the effect of oxygen remaining beyond the tissue consumption, the reason for the increase in oxygen-rich venous flow is unclear. Venous flow in the face is associated with brain cooling~\cite{R29}, and the increase in venous flow observed in the absence of wind might be related to this phenomenon. \par
The rapid changes in $\delta$s shown in Fig.~\ref{fig:Fig6}(a) may be due to vasomotion that compensates for changes in blood pressure and maintains blood flow, that is, the so-called myogenic response~\cite{R30}. The myogenic, metabolic ($PaO_2$, etc.), and neurological controls may have mixed with vasomotion in this study, and further models are expected to be introduced in the future. Accordingly, to identify the microcirculatory structure and quantify hemodynamics, employing a reproduction technique using the NIRS process and windkessel optimization in combination with green light is effective.

\section{Conclusion}

We proposed a vasomotor quantification method using PPG, which has an MA-suppressing structure with eyeglasses and utilizes three wavelengths, including green light. In the cycle ergometer test with and without wind on the face, the integrity of $DC$ and $AC$ signals was improved, and the results demonstrated that only $\Delta [tHb]_{R, IR}$ decreased significantly with wind, whereas both $\Delta [tHb]_{R, IR}$ and $\Delta [tHb]_G$ increased without wind. A microcirculation model, including AVAs estimated from the characteristics in the depth direction, explained this phenomenon and the measured $PWA$s. Both $\delta_A$ and $\delta_{AVA}$ values obtained from the model suggest vasoconstriction with wind and vasodilation after cycling without wind. The vasomotion quantification method developed herein can be effective for tracing autonomic control in real life.


\begin{thebibliography}{99}

\bibitem{R1} 
K. Kotani, {\it et al.}: ``Model for complex heart rate dynamics in health and diseases,'' Phys. Rev. E {\bf 72} (2005) 41904 (DOI: 10.1103/PhysRevE.72.041904).

\bibitem{R2}
K. Nakamura: ``Central circuitries for body temperature regulation and fever,'' Am. J. Physiol. Regul., Integr. Comp. Physiol. {\bf 301} (2011) 1207 (DOI: 10.1152/ajpregu.00109.2011).

\bibitem{R3}
J. B. Mandeville, {\it et al.}: ``Evidence of a cerebrovascular postarteriole windkessel with delayed compliance,'' J. Cereb. Blood. Flow. Metab. {\bf 19} (1999) 679 (DOI: 10.1097/00004647-199906000-00012).

\bibitem{R4} 
M. Xia: ``Tissue and vascular oxygenation dynamics determined by optical approaches and MRI,'' Ph.D Dissertation, The University Texas, Arlington (2007).

\bibitem{R5} 
G. Schumacher, {\it et al.}: ``Multiple coupled resonances in the human vascular tree: refining the westerhof model of the arterial system,'' J. Appl. Physiol. {\bf 124} (2018) 131 (DOI: 10.1152/japplphysiol.00405.2017).

\bibitem{R6} 
T. Ray, {\it et al.}: ``Bio-integrated wearable systems: A comprehensive review,'' Chem. Rev. {\bf 119} (2019) 5461 (DOI: 10.1021/acs.chemrev.8b00573).

\bibitem{R7}
J. Kim, {\it et al.}: ``Wearable biosensors for healthcare monitoring,'' Nat. Biotechnol. {\bf 37} (2019) 389 (DOI: 10.1038/s41587-019-0045-y).

\bibitem{R8} 
T. Y. Abay and P. A. Kyriacou: ``Photoplethysmography for blood volumes and oxygenation changes during intermittent vascular occlusions,'' J. Clin. Monit. Comput. {\bf 32} (2018) 447 (DOI: 10.1007/s10877-017-0030-2).

\bibitem{R9} 
T. Abay and P. Kyriacou: ``Reflectance photoplethysmography as noninvasive monitoring of tissue blood perfusion,'' IEEE Trans. Biomed. Eng. {\bf 62} (2015) 2187 (DOI: 10.1109/TBME.2015.2417863).

\bibitem{R10} 
T. Y. Abay: ``Reflectance photoplethysmography for non-invasive monitoring of tissue perfusion,'' Ph.D Dissertation, City University London (2016).

\bibitem{R11}
S. Segal: ``Regulation of blood flow in the microcirculation,'' Microcirculation {\bf 12} (2005) 33 (DOI: 10.1080/10739680590895028).

\bibitem{R12}
Y. Maeda, {\it et al.}: ``The advantages of wearable green reflected photoplethysmography,'' J. Med. Syst. {\bf 35} (2011) 829 (DOI: 10.1007/s10916-010-9506-z).

\bibitem{R13} 
N. Charkoudian: ``Skin blood flow in adult human thermoregulation: how it works, when it does not, and why,'' Mayo Clin. Proc. {\bf 78} (2003) 603 (DOI: 10.4065/78.5.603).

\bibitem{R14} 
E. Arens and H. Zhang: ``The skin’s role in human thermoregulation and comfort'' in: {\it Thermal and moisture transport in fibrous materials} ed. N. Pan and P. Gibson, (Woodhead Publishing, Cambridge, 2006) 560–602 (DOI: 10.1533/9781845692261.3.560).

\bibitem{R15}
M. M. Prichard and P. M. Daniel: ``Arteriovenous anastomoses in the human external ear,'' J. Anat. {\bf 90} (1956) 309.

\bibitem{R16}
E. A. Tansey, {\it et al.}: ``The sympathetic release test: a test used to assess thermoregulation and autonomic control of blood flow,'' Adv. Physiol. Educ. {\bf 38} (2014) 87 (DOI: 10.1152/advan.00095.2013).

\bibitem{R17}
M. Poh, {\it et al.}: ``Motion-tolerant magnetic earring sensor and wireless earpiece for wearable photoplethysmography,'' IEEE Trans. Inf. Technol. Biomed. {\bf 14} (2010) 786 (DOI: 10.1109/TITB.2010.2042607).

\bibitem{R18}
Y. Ye, {\it et al.}: ``Combining nonlinear adaptive filtering and signal decomposition for motion artifact removal in wearable photoplethysmography,'' IEEE Sens. J. {\bf 16} (2016) 7133 (DOI: 10.1109/JSEN.2016.2597265).

\bibitem{R19}
Y. Zhang, {\it et al.}: ``Motion artifact reduction for wrist-worn photoplethysmograph sensors based on different wavelengths,'' Sensors {\bf 19} (2019) E673 (DOI: 10.3390/s19030673).

\bibitem{R20}
M. Yücel, {\it et al.}: ``Targeted principle component analysis: A new motion artifact correction approach for near-infrared spectroscopy,'' J. Innov. Opt. Health. Sci. {\bf 7} (2014) 1350066 (DOI: 10.1142/S1793545813500661).

\bibitem{R21}
H. Tanaka: ``Instantaneous orthostatic hypotension, postural tachycardia syndrome and neurally mediated syncope in children,'' Acta cardiologica paediatrica Japonica {\bf 17} (2001) 8.

\bibitem{R22}
H. Tanaka, {\it et al.}: ``Cardiac output and blood pressure during active and passive standing,'' Clin. Physiol. {\bf 16} (1996) 157 (DOI: 10.1111/j.1475-097X.1996.tb00565.x).

\bibitem{R23}
M. Laughlin, {\it et al.}: ``Peripheral circulation,'' Compr. Physiol. {\bf 2} (2012) 321 (DOI: 10.1002/cphy.c100048).

\bibitem{R24}
M. Spranger, {\it et al.}: ``Blood flow restriction training and the exercise pressor reflex: a call for concern,'' Am. J. Physiol. Heart Circ. Physiol. {\bf 309} (2015) H1440 (DOI: 10.1152/ajpheart.00208.2015).

\bibitem{R25}
A. Reisner, {\it et al.}: ``Utility of the photoplethysmogram in circulatory monitoring,'' Anesthesiology {\bf 108} (2008) 950 (DOI: 10.1097/ALN.0b013e31816c89e1).

\bibitem{R26}
J. Coffman and A. Cohen: ``Total and capillary fingertip blood flow in raynaud's phenomenon,'' N. Engl. J. Med. {\bf 285} (1971) 259 (DOI: 10.1056/NEJM197107292850505).

\bibitem{R27}
M. Shibata, {\it et al.}: ``Microvascular and interstitial $\rm P_{O_2}$ measurements in rat skeletal muscle by phosphorescence quenching,'' J. Appl. Physiol. {\bf 91} (2001) 321 (DOI: 10.1152/jappl.2001.91.1.321).

\bibitem{R28}
T. Sakai and Y. Hosoyamada: ``Are the precapillary sphincters and metarterioles universal components of the microcirculation? An historical review,'' J. Physiol. Sci. {\bf 63} (2013) 319 (DOI: 10.1007/s12576-013-0274-7).

\bibitem{R29}
K. Dohi, {\it et al.}: ``Positive selective brain cooling method: a novel, simple, and selective nasopharyngeal brain cooling method,'' Acta. Neurochir. Suppl. {\bf 96} (2006) 409 (DOI: 10.1007/3-211-30714-1\_84).

\bibitem{R30}
R. Berg, {\it et al.}: ``Static cerebral blood flow autoregulation in humans,'' Curr. Hypertens. Rev. {\bf 5} (2009) 140 (DOI: 10.2174/157340209788166922).

\end{thebibliography}
\end{document}